# Towards an Electronic Health Record System in Vietnam: A Core Readiness Assessment

Stefan HOCHWARTER[a,1], Do Duy CUONG[b], Nguyen Thi Kim CHUC[c] and Mattias Larsson[a]

[a] *Karolinska Institutet, Department of Public Health Sciences, Stockholm, Sweden*
[b] *Bach Mai Hospital, Infectious Disease Department, Hanoi, Vietnam*
[c] *Hanoi Medical University, Hanoi, Vietnam*

**Abstract.** Previous studies have shown that health information technologies have a positive impact on health systems. Electronic health record (EHR) systems are one of the most promising applications, demonstrating a positive effect in high income countries. On the other hand, robust evidence for low and middle income countries is still sparse. The aim of this study is to initiate a carefully planned nationwide EHR system in Vietnam by assessing the core readiness.

The assessment structure is mainly based on previous research, which recommends a readiness assessment prior to an EHR system implementation. To collect data, participant observation, document analysis and an in-depth interview were used.

This study has revealed new insights into the current situation on EHR in Vietnam. The Ministry of Health is currently working on improving the conditions for future implementation of a Vietnamese EHR system. There are issues with the current way of handling health records. These issues are encouraging the Ministry of Health to work on identifying the next steps for an EHR system implementation. The integration of an EHR system with current systems seems to be challenging, as most systems are commercial, closed source and very likely have no standardised interface/gateway.

In conclusion, this study identifies points which need to be further investigated prior to an implementation. Generally, health care workers show good awareness of new technologies. As the Vietnam's health care system is centrally organised, there is the possibility for a nation-wide implementation. This could have a great positive impact on the health care system, however, besides rigours planning also standards need to be followed and common gateways implemented. Finally, this assessment has focused only on one level of a readiness assessment. Further research is needed to complete the assessment.

## 1. Introduction

Health information technologies (HIT) show promising positive effects on a nation's health systems in terms of efficiency, safety and effectiveness. A significant improvement on patient safety can be reached by using Computerized

---
[1] Corresponding Author: stefan.hochwarter@gmail.com

Physician Order Entry (CPOE) systems, both in inpatient and ambulatory settings. Furthermore, an Electronic Medical Record (EMR) system has the potential to contribute to a healthier community by disease prevention measures and chronic disease management. [1]

Well implemented EMR systems or respectively Electronic Health Record (EHR) systems can be seen as the foundation for positive outcomes of HIT systems. This article focuses on EHR systems in developing countries by investigating the current situation in Vietnam. Therefore, at this point a definition for EMR and EHR is necessary. EHR can be defined as follows.

*The EHR means a repository of patient data in digital form, stored and exchanged securely, and accessible by multiple authorized users. It contains retrospective, concurrent, and prospective information and its primary purpose is to support continuing, efficient and quality integrated health care.* [2]

The above stated definition is based on the International Organization for Standardization (ISO). The same paper invested different types and terms of EHR by conducting a literature review. The majority of papers using EMR are referring to a "Departmental EMR", meaning that it is "generally focused on medical care" and that information is "entered by a single hospital department". [2] Also, these definitions of EMR and EHR are similar to those in the World Health Organization's (WHO) "Electronic Health Records - Manual for Developing Countries", which was referred to during this research. [3]

Benefits of using EHR/EMR systems have been proven in studies for high income countries. However, the same robust evidence for low and middle income countries (LMIC) is still lacking despite the several success stories of implementations. [4] As more and more high income countries proceed to use HIT and EHR/EMR systems in particular, pitfalls, challenges and barriers were identified. Most named pitfalls include "lack of user training; poor initial design […]; systems difficult to use or complex; dependence on one individual 'champion'; lack of involvement of local staff in design and testing; lack of perceived benefit for users who collect data." [5] Many challenges were of technical nature, related to low computer literacy, poor infrastructure, insufficient back-up systems (in terms of data as well as computer-loss), poor system security, lack of resources, insufficient political commitment and support. [5] [6] Moreover, local constraints such as "political, social, cultural" ones need to be brought to mind. [7] [8] Due to the higher complexity of HIT projects in LMIC (in comparison to high income countries, where many stakeholder are already experienced in HIT projects and more previous research exists), continuous evaluation and considerate planning for all stakeholders and local circumstances are crucial for success. [9] [4]

In many high income countries the EHR/EMR systems where initially developed at hospital level and in some cases even on department level, hence why there are many competing and badly integrated systems. This leads to additional costs in development and maintenance of the systems. For the health staff multiple systems lead to poor access to patient records and time consuming procedures to get adequate patient information such as previous health care seeking, examinations and tests results. This in turn leads unnecessary health care consumption including multiple similar examinations and laboratory tests.

For LMIC it might be an advantage to learn from the experience in high income countries and coordinate the EHR/EMR system implementation towards more integrated systems.

In Vietnam, a country with 90 million inhabitants on the brink of becoming a middle income country, EHR/EMR systems have been. This has in most cases been driven more by the need for accounting rather than the patient care. Previous studies have shown that integration of all stakeholders of different categories, such as the Ministry of Health (MoH) is a requirement to successfully finish HIT projects in LMIC as Vietnam. Previous studies showed that integration of all stakeholders of different categories, such as the Ministry of Health (MoH) is a requirement to successfully finish HIT projects in Vietnam. Tran Dac Phu et.al reported that they succeeded in rolling out an electronic communicable disease reporting system in 38 of the 63 Vietnamese provinces and that they expect to have this system in all 63 provinces by 2014. [10] One reason for this great accomplishment might be the increased "share of government health spending at the local level (provincial level and below)." [11] Also, healthcare workers are open-minded in terms of innovative new technologies in the application of health care. However, previous researchers have stressed the importance of careful planning. [10] [12]

The aim of this study is to take the first steps toward an EHR system in Vietnam by assessing the core readiness of the current Vietnamese health care system. The term "core readiness" will be further defined in the Method section and is based on different recommendations and frameworks. Based on the results, new findings and a recommendation for future work will be discussed. This paper is structured similar as recommended in the "Statement on reporting of evaluation studies in Health Informatics". [13]

When implementing an EHR system, there are many ethical points to consider. In LMIC there are several layers to be aware of, in addition to the critical discussion items faced by high income countries, such as "privacy, confidentiality, data security, informed consent, data ownership, and secondary use of data". Those previously mentioned points are already broadly discussed and certainly need to be addressed prior to an implementation. In LMIC, the scope of an EHR implementation should be assessed in the current environment. Resources are normally limited and therefore an evaluation on the "risk-benefit ratio" was recommended. Perhaps there could be other interventions which would have a greater impact on the health system. Were et. al. presented a set of points for EHR systems in developing countries, which need to be considered additional to a readiness assessment. [14]

## 2. Methodology

In this section, first the study context will be presented followed by a detailed description on the study design.

*2.1 Study context*

As the aim of this study is to investigate the core readiness of the Vietnamese health care system to introduce an EHR system, the study was naturally conducted in Vietnam. Table 1 shows key data about Vietnam and its health care system. Notably almost 70% of the Vietnamese are living in rural areas, which geographically is a challenge for health service delivery.

| **Table 1:** Key data about Vietnam and the Vietnamese health care system [15][16][17] | |
|---|---|
| Population 2011 (in millions) | 87.8 |
| Surface area 2011 (in thousand sq. km) | 331.1 |
| Population density 2011 (people per sq.km) | 283 |
| Urban population 2011 (% of total population) | 31 |
| Gross national income per capita 2011 (in USD) | 1 270 |
| Gross domestic product per capita 2010-11 (% growth) | 4.8 |
| Population below USD 1.25 a day 2008 (in %) | 16.9 |
| Prevalence of child malnutrition underweight 2005-11 (% of children under age 5) | 20.2 |
| Youth literacy rate 2005-11 (% ages 15-24) | 97 |
| Mobile cellular subscriptions 2011 (per 100 people) | 143 |
| Individuals using the Internet (% of population) | 35 |
| Life expectancy at birth 2011 (in years) | 75 |
| Total health expenditure 2010 (% of GDP) | 6.9 |

Figure 1 shows the different levels of health service delivery in Vietnam. The referral path is normally bottom-up, so the commune level delivers most primary care services whereas on national level hospitals include wards highly specialised such as Oncology. Besides those levels of service illustrated in Figure 1, there is also a private sector with 102 hospitals and clinics. [17]

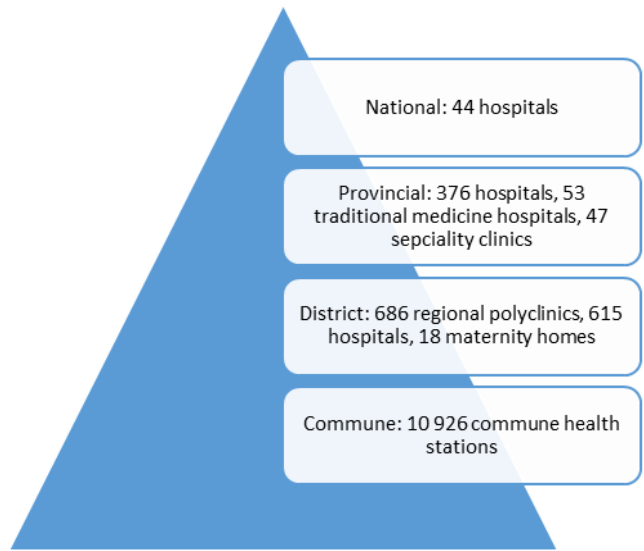

**Figure 1.** Public curative care services in Vietnam [17].

*2.2 Study design*

Based on previous research, a list of questions were defined to assess the core readiness of the Vietnamese health system for introducing an EHR system. Due to limited previous research, the study had to explore many different aspects and as a result the study uses a qualitative research design. Following different frameworks, defining the core readiness, assessing Vietnam and its health care system as a whole, this study is of deductive nature with a holistic perspective.

**Listing 1.** Question catalogue for core readiness assessment

1. Identification of needs for future changes, which the proposed EHR system will address (Interview with Ministry of Health)
2. Dissatisfaction with status quo on the prioritised needs related to the EHR (mixed):
   a. Are medical records currently kept on all patients - inpatients, outpatients and accident and emergency patients?
   b. What type of medical record is kept? Are medical records well documented? What is the quality of the medical record? Has all essential information been recorded, are all entries signed and dated? Are quality checks performed on current paper records? If so, have any documentation problems been identified?
   c. Is the medical record system centralised using a unit numbering system? That is, are all admissions, outpatient notes and accident and emergency records filed under one number in the one medical record?
   d. How are patients identified? Do all persons have a national identification number? Is this used to uniquely identify the patient? If a national identification number is not issued what information is used to identify each patient?
   e. Are daily admissions and discharge lists produced?
   f. Are medical records returned to the medical record department on discharge of the patient?
   g. How are medical records filed?
   h. Is there a problem with duplicate medical records?
   i. When the medical record has been completed by the doctor do medical record staff code the main condition using a classification system such as ICD-10? Do they code procedures? Are the codes indexed to enable the retrieval of medical records for research, health statistics and epidemiological studies?
3. Awareness about EHR systems in the institutes? (answered by previous research)
4. Trust on the use of ICT. (answered by previous research)
   a. Do you think that ICT can address the current problems with paper-based records?
   b. Do you have any concerns when using ICT in a health care setting?
5. Planning for the new EHR project (Interview MoH)
6. Integration of technology (Interview and observation)
   a. Has the integration with current services been considered in the planning process?
   b. Is there a plan in place to integrate the EHR system with current HIS in use.
   c. Use of HI standards

Gateways/interfaces

To collect data, participant observation, document analysis and an in-depth interview were used. Participant observation and document analysis was conducted in the Infectious Disease Department of the Bach Mai Hospital in Hanoi. The participant observation (five visits including one introductive meeting) was carried out over 6 weeks in November and December 2013. Field notes were also made. The documents (such as health records) were analysed together at the Bach Mai Hospital in Hanoi in December 2013. The in-depth interview (60 minutes) was performed in Hanoi with an administrative employee of the Ministry of Health responsible for an eventual nationwide EHR system in Vietnam. Listing 1 shows the resulting questions and states the techniques used to capture data. Certain questions were not followed up as they already had been answered in previous research. This is also indicated in listing 1.Previous research

For the sake of completeness, previous research for question from Listing 1 will be shortly stated in the following paragraphs.

- Awareness about EHR systems in the institutes?
  "*Healthcare workers at all levels show great interest in computerization and networking of the public healthcare system. They are aware of the relevant application for health care.*" [12]
  As Trân et.al. state in 2004 in their research, healthcare workers in Vietnam are well aware of the existence and benefits of an EHR system.

- **Trust on the use of ICT**
  "*Growing technology maturity and increased training options, the government is ready to invest in and aware of technologies*" [25] Shilaaber recent research shows that healthcare workers in Vietnam generally are willing to use new technologies.

Applying e-health readiness assessments in developing countries is not a new idea and research was done with different perspectives.[24] The WHO stresses the importance of detailed planning and points out that an implementation of an EHR system should only start after a readiness assessment. [3] There are different areas and aspects of e-health readiness. [18] E-health can be further categorised into core readiness, technological readiness, learning readiness, societal readiness and policy readiness. Core readiness deals with the "overall planning process for a proposed e-health program, and the knowledge and experience of planners with programs using ICT". [19] A conceptual framework was developed following those categories with focus on developing countries. [20] Consequently this framework was successfully tested in Pakistan while assessing its e-Health readiness. [21] Also the framework was backed up by another article, which made an interesting approach to quantify the assessment using a graph-based approach. [22] Listing 1 was based on the previous research mentioned in this paragraph.

The data collection was performed in November and December 2013. One interview was performed at the Ministry of Health in Hanoi and the observation and document analysis took place at the Bach Mai Hospital Hanoi.

## 3. Results

In this section the results will be presented, following the same structure as defined in Listing 1.

*3.1 Identification of needs for future changes, which the proposed EHR system will address*

The Ministry of Health is currently working on creating the appropriate conditions that are necessary to introduce an EHR system in Vietnam. For example, it was identified that a new patient record number needs to be introduced, which will be used throughout the Vietnamese health care system. [23] (Reference received from MoH.) Also they are well aware of the technical challenges they will face, considering the four levels of the health care system and the fact that 70% of the population is living in rural areas.

*3.2 Dissatisfaction with status quo on the prioritised needs related to the HER*

(a) Medical records are currently kept on all patients, which is (most of the time) a paper-based record.
(b) The structure of the paper-based medical records is defined by the Ministry of Health. The quality of the medical records varies between hospitals and clinics. Every two weeks there is a follow up on the medical records and there are random controls every six months.

(c) The medical record system is not centralised and so, there may be multiple medical records of one patient in different wards with different identification numbers.

(d) People have a unique insurance number, which could be used for numbering but as this number often is not known, a ward is usually using its own numbering system. So in theory there would be a national identification number but as patients often do not know their own national identification number and might also miss an insurance card, the identification number is not used for primary identification in (E)HRs.

(e) It is common practice, that daily admissions and discharge lists are produced.

(f) Due to the fact that different identification numbers and multiple records in different wards, there are also problems with duplicates.

(g) There is also an archive system for the medical records, they return to the medical record department on discharge of the patient and are stored for 10 years, or for 20 years in case of death.

(h) The medical records are stored in filed lockers and sorted by the patient record number.

(i) Physicians and nurses are aware of ICD-10 classification, but it is used for research purposes only.

*3.3 Planning for the new EHR project*

At the Ministry of Health (MoH), there is a strong intention to implement an EHR system. However, the MoH is aware that they do not fulfil the requirements for an implementation yet. They investigate the current situation and try to identify the next steps, for example introducing a new patient record number. As there is no concrete implementation plan yet, there is no project plan including budgets and time plan.

*3.4 Integration of technology*

(a) As mentioned above, there is no concrete project plan for an EHR implementation yet. The integration of current services is desirable but may be hard to realise due to the high variety within different service providers.

(b) Current HIS in use can be found related to billing of health expenses. Again, according to the MoH it is desirable to integrate existing systems with the EHR system but as those systems often lack standards and / or gateways, this will be rather a challenge.

(c) There are some private vendors offering EMR systems. They claim to follow the standard Health Level Seven (HL7) but this is hard to verify as they are closed source and mostly only used as an isolated system.

(d) Furthermore, there are no experiences in exchanging data using well-defined interfaces or gateways.

## 4. Discussion

Even though the MoH has taken initial steps towards an EHR system implementation, there are still many things to improve to reach core readiness for an EHR implementation. As reported, Vietnam plans to introduce a common unique patient identifier in the near future. This is one core requirement for successfully introducing EHRs. However, this will be a complex step as the Vietnamese health system with its four levels of health service delivery and rural population is a challenging environment. On the other hand, the strong MoH and the good compliance in lower hierarchies can back this ambitious undertaking.

Health care workers are well aware of the advantages of using an EHR but also express their concerns regarding the risks, such as violation of patient's privacy. Thus, it needs to be stressed once again that prior to an implementation, a considerate, detailed planning stage including a complete readiness assessment is crucial. This will point out risks, challenges and give the opportunity to address and provide strategies to overcome them.

Due to the fact that there will most likely not be a single vendor offering all the EHR systems for all levels of health service delivery and all regions in Vietnam, the assessment on integration of technology plays an important role in this assessment. To guarantee that the systems can work together but at the same time be designed for the specific requirements of the respective users, standards need to be followed and gateways need to be defined for communication between different systems. [26] [27]

As the health care system in Vietnam is relatively centralised under MoH and the provincial health authorities, there may be a possibility to achieve a nationwide integrated EHR/EMR systems, at least in the 44 central hospitals and possibly also at provincial level. This could in such case be one of the larger integrated EHR/EMR systems worldwide. If this could be attained it would be of interest to many LMIC, in addition health care system benefits could be assessed. The assessment would be in relation to referrals, savings due to access of data and decrease of multiple examinations and laboratory tastings. It could also be an effective way of canalising competence from tertiary to provincial and district level health care, for example updating easy to access treatment guidelines through the EHR/EMR systems.

This study has only assessed one part of e-health readiness and further research is required to get a complete picture of the current situation. A new study could investigate the different perception of health care providers (such as physicians, health community workers) and health care managers (such as policy makers in the MoH). Also comparative research between the different health sectors may be of interest. Certainly, the other categories of e-health readiness (for example technological readiness or societal readiness) must be assessed prior to an implementation. Finally, similar as done in this study, using one or more frameworks, a complete tool for readiness assessment for an EHR system in Vietnam can be implemented and consequently evaluated and applied.

# References

[1] Hillestad, R., Bigelow, J., Bower, A., Girosi, F., Meili, R., Scoville, R., & Taylor, R. (2005). Can electronic medical record systems transform health care? Potential health benefits, savings, and costs. *Health Affairs (Project Hope)*, **24**(5), 1103–17. doi:10.1377/hlthaff.24.5.1103

[2] Häyrinen, K., Saranto, K., & Nykänen, P. (2008). Definition, structure, content, use and impacts of electronic health records: a review of the research literature. *International Journal of Medical Informatics*, **77**(5), 291–304. doi:10.1016/j.ijmedinf.2007.09.001

[3] World Health Organization. (2011). Electronic Health Records - Manual for Developing Countries. *Vasa*.

[4] Hersh, W., Margolis, A., Quirós, F., & Otero, P. (2010). Building a health informatics workforce in developing countries. *Health Affairs (Project Hope)*, **29**(2), 274–7. doi:10.1377/hlthaff.2009.0883

[5] Sood, S. P., Nwabueze, S. N., Mbarika, V. W. A., Prakash, N., Chatterjee, S., Ray, P., & Mishra, S. (2008). Electronic Medical Records: A Review Comparing the Challenges in Developed and Developing Countries. *Proceedings of the 41st Annual Hawaii International Conference on System Sciences (HICSS 2008)*. doi:10.1109/HICSS.2008.141

[6] Bukachi, F., & Pakenham-Walsh, N. (2007). Information technology for health in developing countries. *Chest*, **132**, 1624–1630. doi:10.1378/chest.07-1760

[7] Burney, A., Mahmood, N., & Abbas, Z. (2010). Information and communication technology in healthcare management systems: prospects for developing countries. *International Journal of Computer Application*, **4**(2), 27–32.

[8] HASANAIN, R. A., & COOPER, H. (2014). Solutions to Overcome Technical and Social Barriers to Electronic Health Records Implementation in Saudi Public and Private Hospitals. *Journal of Health Informatics in Developing Countries*, **8**(1).

[9] Fraser, H. S., & Blaya, J. (2010). Implementing medical information systems in developing countries, what works and what doesn't. *AMIA ... Annual Symposium Proceedings / AMIA Symposium. AMIA Symposium*, *2010*, 232–6. Retrieved from http://www.pubmedcentral.nih.gov/articlerender.fcgi?artid=3041413&tool=pmcentrez&rendertype=abstract

[10] Tran, P. D., Vu, L. N., Nguyen, H. T., Phan, L. T., Lowe, W., McConnell, M. S., … Kinkade, C. (2014). Strengthening global health security capacity - Vietnam demonstration project, 2013. *MMWR. Morbidity and Mortality Weekly Report*, **63**(4), 77–80. Retrieved from http://www.ncbi.nlm.nih.gov/pubmed/24476979.

[11] Prepared, P., Symposium, I., & Systems, H. C. (2005). *Vietnam's Health Care System : A Macroeconomic Perspective*.

[12] Trân, V. A., Seldon, H. L., Hoàng, D. C., & Nguyên, K. P. (2006). Electronic healthcare communications in Vietnam in 2004. *International Journal of Medical Informatics*, **75**(10-11), 764–70. doi:10.1016/j.ijmedinf.2006.01.002.

[13] Talmon, J., Ammenwerth, E., Brender, J., de Keizer, N., Nykänen, P., & Rigby, M. (2009). STARE-HI--Statement on reporting of evaluation studies in Health Informatics. *International Journal of Medical Informatics*, **78**(1), 1–9. doi:10.1016/j.ijmedinf.2008.09.002.

[14] Were, M. C., & Meslin, E. M. (2011). Ethics of implementing Electronic Health Records in developing countries: points to consider. *AMIA ... Annual Symposium Proceedings / AMIA Symposium. AMIA Symposium*, *2011*, 1499–505. Retrieved from http://www.pubmedcentral.nih.gov/articlerender.fcgi?artid=3243215&tool=pmcentrez&rendertype=abstract.

[15] The World Bank. (2013). *World development indicators*.

[16] World Health Organization. (2011). *Viet Nam: Health Profile*.

[17] World Health Organization, & Ministry of Health Viet Nam. (2012). *Health Service Delivery Profile Viet Nam*.

[18] Ajami, S., Ketabi, S., Isfahani, S., & Heidari, A. (2011). Readiness assessment of electronic health records implementation. *Acta Informatica Medica*, **19**(November), 224–227. doi:10.5455/aim.2011.19.224-227.

[19] Khoja, S., Scott, R. E., Casebeer, A. L., Mohsin, M., Ishaq, a F. M., & Gilani, S. (2007). e-Health readiness assessment tools for healthcare institutions in developing countries. *Telemedicine Journal and E-Health : The Official Journal of the American Telemedicine Association*, **13**(4), 425–31. doi:10.1089/tmj.2006.0064.

[20] Khoja, S., Scott, R., Mohsin, M., Ishaq, A., & Casebeer, A. (2007). Developing a conceptual-framework for e-health readiness assessment tools for developing countries. *International Hospital ...*, 2007–2009.

[21] Khoja, S., Scott, R., & Gilani, S. (2008). E-health readiness assessment: promoting "hope" in the health-care institutions of Pakistan. *World Hospitals and Health Services : The Official Journal of the International Hospital Federation*, **44**(1), 36–8. Retrieved from http://www.ncbi.nlm.nih.gov/pubmed/18549033

[22] Li, J., Land, L. P. W., Ray, P., & Chattopadhyaya, S. (2010). E-Health readiness framework from Electronic Health Records perspective. *International Journal of Internet and Enterprise Management*. doi:10.1504/IJIEM.2010.035626.

[23] Thanh Nien. (2013). *Đẩy nhanh tiến độ cấp mã số công dân*. Retrieved March 10, 2014, from http://www.thanhnien.com.vn/pages/20131207/day-nhanh-tien-do-cap-ma-so-cong-dan.aspx.

[24] Adjorlolo, S., & Ellingsen, G. (2013). Readiness Assessment for Implementation of Electronic Patient Record in Ghana: A Case of University of Ghana Hospital. *Journal of Health Informatics in Developing Countries*, **7**(2).

[25] Shillabeer, A., & Anderson, B. (2012). *The Public Health Context in Vietnam and the Potential for Informatic*; notes provided at the congress MEDINFO 2013. Retrieved from http://person.hst.aau.dk/ska/MEDINFO2013/AllPresentations/Pdfs/Symposiums/C4_942_MEDINFO2013.pdf.

[26] Braa, J., Hanseth, O., & Heywood, A. (2007). Developing Health Information Systems in Developing Countries: The Flexible Standards Strategy. *Mis Quarterly*, **31**(August), 1–22. Retrieved from http://www.jstor.org/stable/25148796.

[27] Saleem, T. (2009). Implementation of EHR/EPR in England: a model for developing countries. *Journal of Health Informatics in Developing Countries*, **3**(1).